\documentclass[twoside]{ilcws10}
\usepackage[latin1]{inputenc}
\usepackage[dvips]{graphicx,epsfig,color}
\usepackage{wrapfig,rotating}
\usepackage{amssymb,amsmath,array}

\begin{document}
\title{
Impacts of SB2009 on the Higgs Recoil Mass Measurement Based on a Fast Simulation Algorithm for the ILD Detector} 

 \author{Hengne LI \\
\vspace{.3cm}\\
Laboratoire de Physique Subatomique et de Cosmologie (LPSC),\\
38026 Grenoble Cedex, France\\
\vspace{.1cm}\\
Laboratoire de l Acc\'el\'erateur Lin\'eaire (LAL),   \\
91898 Orsay Cedex, France \\
}

\maketitle

\begin{abstract}
This proceeding reports a study of the impacts of the SB2009 beam parameters on the Higgs recoil mass and Higgs-Strahlung cross-section measurements based on a dedicated fast simulation algorithm of the ILD detector. The study shows worse results from SB2009 beam parameters than the previous RDR beam parameters, because of the smaller luminosity. However, Travel Focus (TF) technology can recover the degradation to certain level. 
\end{abstract}

\section{Introduction}

The intent of this work is to study of the impacts of the SB2009~\cite{beampar} beam parameters on the Higgs recoil mass and Higgs-Strahlung cross-section measurements based on a dedicated fast simulation algorithm of the ILD~\cite{ildloi} detector. We would like to compare this result with our previous result~\cite{ildloi, thesis, note} at $\sqrt{s}=$250~GeV with RDR~\cite{rdr} beam parameters (RDR 250) for the accelerator optimization reason. 

The study uses the $ZH\rightarrow\mu^+\mu^-X$ channel, assuming a beam polarization of ($e^-: -80\%,~e^+: +30\%$). It takes into account the beam effects by beam simulation using GUINEA-PIG~\cite{guineapig} with beam parameters SB2009. I use PYTHIA~\cite{pythia} for the event generation with the beams simulated by GUINEA-PIG as inputs through the interface CALYPSO~\cite{calypso}. Thereafter, I developed a fast simulation algorithm~\cite{fastsim} to include the detector effects of the ILD. After the fast simulation, I perform the analysis based on the same algorithm as of my previous study~\cite{thesis, note}. In the end, I give the result and the comparison.

\begin{figure}[htpb]
\centering
\includegraphics[width=0.8\textwidth]{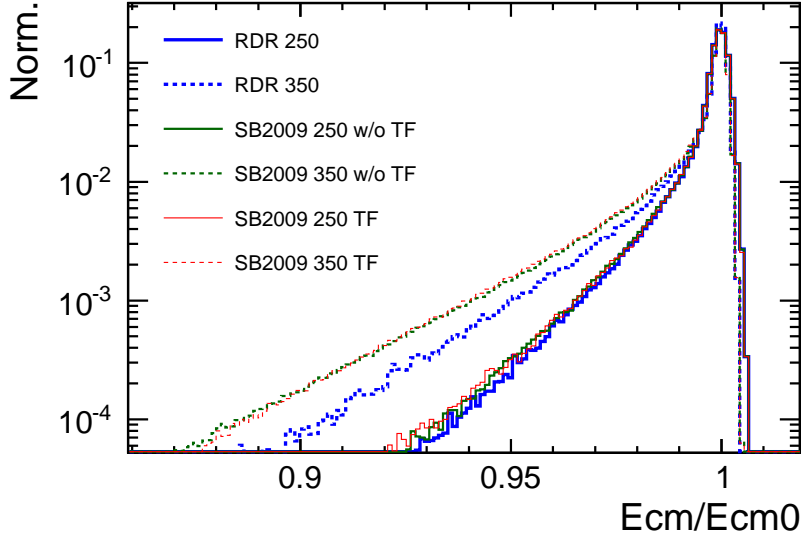}
\caption{Normalized luminosity spectra simulated with beam parameters of SB2009 and RDR.}
\label{fig:lumi}
\end{figure}

\section{Beam Simulation}

The beam effects are simulated using GUINEA-PIG, with the beam parameters SB2009 given by reference~\cite{beampar}. Figure~\ref{fig:lumi} shows the luminosity spectra simulated with beam parameters SB2009, comparing with RDR beam parameters. .

\section{Event Generation}

I use PYTHIA for the event generation. The event generation takes the beams simulated by GUINEA-PIG  as inputs through the interface CALYPSO. 

I take the $ZH\rightarrow\mu^+\mu^-X$ channel under study, with two major background reactions the $WW$ ($W^+W^-\rightarrow\mu^+\nu_{\mu}\mu^-\bar{\nu_{\mu}}$) and the $ZZ$ ($ZZ\rightarrow\mu^+\mu^-f\bar{f}$). Their cross-sections at $\sqrt{s}=$350~GeV with beam polarization ($e^-: -80\%,~e^+: +30\%$) are listed in Table~\ref{tab:xsec}. 

\begin{table}[htb]
\centering
\begin{tabular}{|c|c|}
\hline
Reaction &  Cross-Section \\
\hline
\boldmath{$ZH\rightarrow\mu\mu X$} & \bf{7.1 fb} \\
 \hline
$WW$ & 346 fb \\
$ZZ$ & 165 fb \\
 \hline
\end{tabular}
\caption{Reactions and cross sections at $\sqrt{s}=$350~GeV with beam polarization ($e^-: -80\%,~e^+: +30\%$). The signal is indicated by bold face letters.}
\label{tab:xsec}
\end{table}

If I take the RDR 500 peak luminosity ($\mathcal{L}_{{\rm peak,RDR500}}=2.0\times10^{34}$cm$^{-2}$s$^{-1}$) and integrated luminosity ($\mathcal{L}_{{\rm int,RDR500}}=$500~fb$^{-1}$) as reference, the estimated integrated luminosity of a given set of beam parameters should be~\cite{francois}:
\begin{equation}
\mathcal{L}_{{\rm int}}=\frac{\mathcal{L}_{{\rm peak}}}{\mathcal{L}_{{\rm peak,RDR500}}} \cdot \mathcal{L}_{{\rm int,RDR500}}
\end{equation} 

Following this rule, these integrated luminosities for various beam parameters are listed in Table~\ref{tab:lumi}.

\begin{table}[htb]
\centering
\begin{footnotesize}
\begin{tabular}{|r|l|l|l|l|l|l|l|l|l|l|l|}
\hline
  &\multicolumn{3}{c|}{RDR} & \multicolumn{4}{c|}{SB2009 w/o TF} & \multicolumn{4}{c|}{SB2009 w/ TF} \\
\hline
 $\sqrt{s}$ (GeV) &  250 & 350 & 500 & 250.a & 250.b & 350 & 500 & 250.a & 250.b & 350 & 500 \\
\hline
Peak L ($10^{34}$cm$^{-2}$s$^{-1}$) & 0.75 & 1.2 & 2.0 & 0.2 & 0.22 & 0.7 & 1.5 & 0.25 & 0.27 & 1.0 & 2.0 \\
 \hline
Integrated L (fb$^{-1}$) & 188 & 300 & 500 & 50 & 55 & 175 & 375 & 63 & 68 & 250 & 500 \\
 \hline
\end{tabular}
\end{footnotesize}
\caption{Estimated Integrated luminosities for various beam parameters~\cite{beampar}. }
\label{tab:lumi}
\end{table}

\section{Fast Simulation}

I developed a dedicated fast simulation algorithm for the ILD detector concept. 

The fast simulation of the ILD detector intents to consider the uncertainty of the detector response without full detector simulation. For the Higgs recoil mass analysis, the major detector uncertainty is due to the momentum measurement of the lepton tracks. I take the muons tracks for this study. Therefore the object of the fast simulation is to smear the MC true momentum given by event generation according to the momentum resolution of the detector. It consists of two steps.

The first step is to have the momentum resolution of the ILD detector. I parameterize the momentum resolution of the ILD detector as a function of the momentum ($P$) and $\cos{\theta}$ of muon tracks. The momentum resolution function is given by Equation~\ref{eq:dpop2}.

\begin{equation}
\frac{\Delta P}{P^2} = \left\{ \begin{array}{ll}
a_1 \oplus b_1/P & :\ |\cos{\theta}|<0.78\\
(a_2 \oplus b_2/P)\bigg/\sin{(1-|\cos{\theta}|)} & :\ |\cos{\theta}|>0.78
\end{array} \right. 
\label{eq:dpop2}
\end{equation}

\begin{figure}[htpb]
\centering
\includegraphics[width=0.6\textwidth]{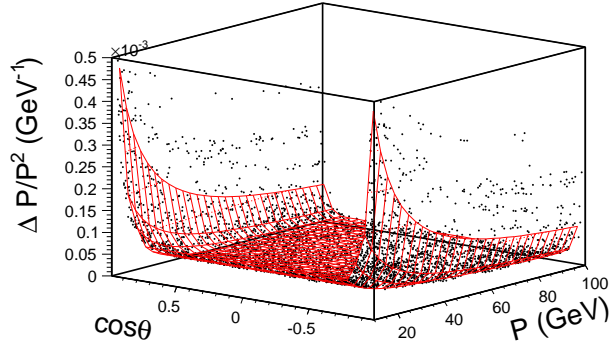}
\caption{The black scatter plot shows the distribution of momentum resolution of the ILD detector as a function of lepton $P$ and $\cos\theta$ obtained from full simulation of the detector. The red surface shows the fit of Equation~\ref{eq:dpop2} to the distribution. The parameters obtained from the fit is shown in Table~\ref{tab:dpop2}. }
\label{fig:dpop2}
\end{figure}

\begin{table}[htdp]
\centering
\begin{tabular}{c|l}
\hline
$a_1$ & $2.08\times10^{-5}$ (1/GeV) \\
\hline
$b_1$ & $8.86\times10^{-4}$  \\
\hline
$a_2$ & $3.16\times10^{-6}$ (1/GeV) \\
\hline
$b_2$ & $2.45\times10^{-4}$  \\
\hline
\end{tabular}
\caption{Parameters in Equation~\ref{eq:dpop2} obtained by fitting it to the distribution of momentum resolution of the ILD detector shown in Figure~\ref{fig:dpop2}.}
\label{tab:dpop2}
\end{table}

Equation~\ref{eq:dpop2} is thus fitted to the distribution of momentum resolution got from the full simulation of the ILD detector, as shown in Figure~\ref{fig:dpop2}. In this figure, the black scatter plot shows the momentum resolution as a function of $P$ and $\cos\theta$ from ILD detector full simulation, and the red surface is the fit of Equation~\ref{eq:dpop2} to the distribution of momentum resolution. The momentum resolution for a given lepton is obtained by propagation of the error matrix of its track reconstruction. The fitted parameters are shown in Table~\ref{tab:dpop2}.

The second step of the fast simulation is to smear the MC true momentum from the event generation according to the momentum resolution function just obtained. For each lepton, I define a Gaussian function, and set the MC true momentum as the mean of this Gaussian function, the $\Delta P$ as the sigma. Thereafter, I generate a random number according to this Gaussian function, and this random number is the momentum after fast simulation.

\begin{figure}[htpb]
\centering
\includegraphics[width=0.5\textwidth]{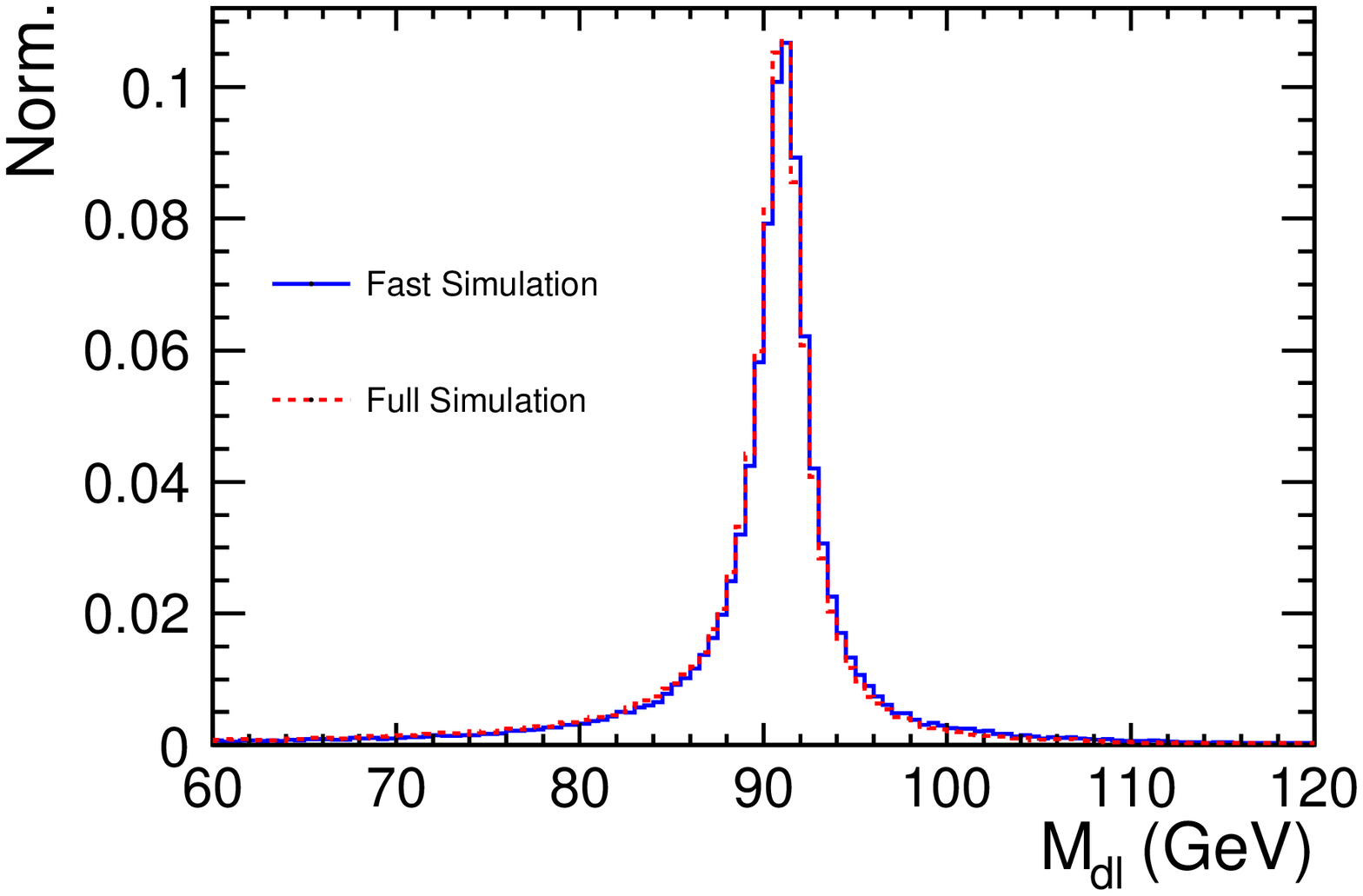}%
\includegraphics[width=0.5\textwidth]{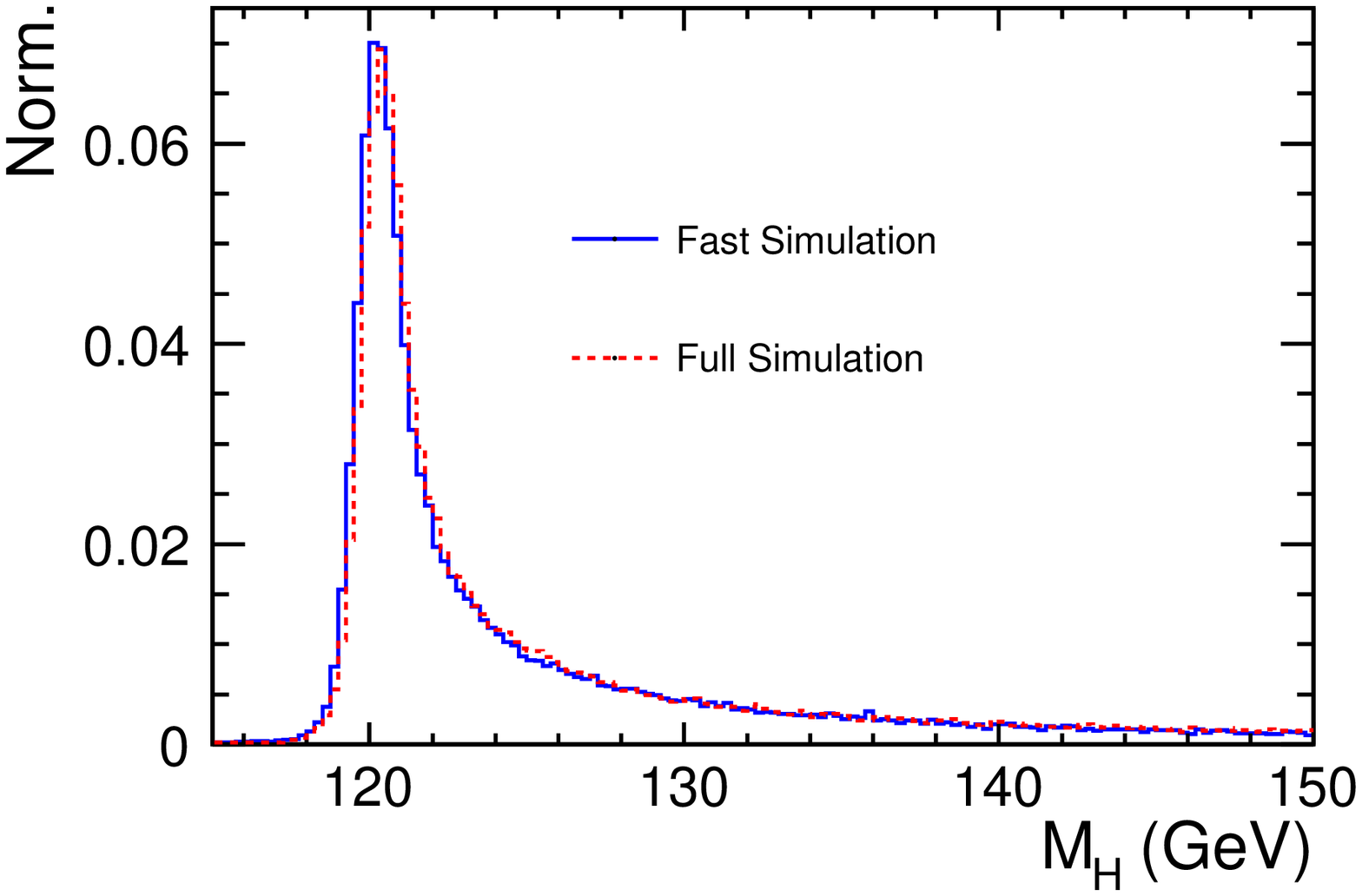}
\caption{Comparison of the invariant mass $M_{dl}$ of the lepton pair (top) and the recoil mass $M_H$ (bottom) distributions from fast simulation and full simulation of the ILD detector at $\sqrt{s}=$250~GeV.  }
\label{fig:mdl_mh_rdr250_fs}
\end{figure}

\begin{figure}[htbp]
\centering
\includegraphics[width=0.6\textwidth]{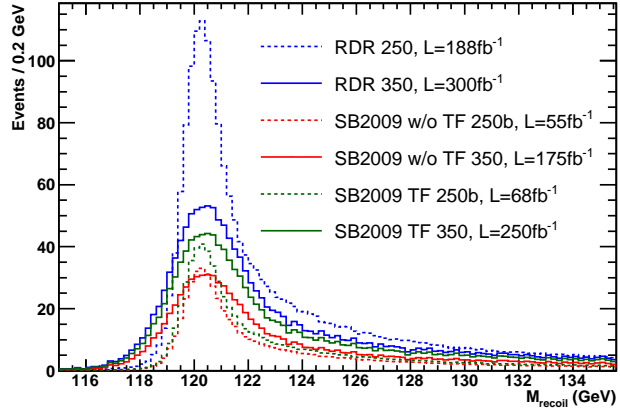}
\caption{Comparison of the Higgs recoil mass distributions of the signal after fast simulation with various beam parameters considering the corresponding integrated luminosity. The beam polarization is ($e^-: -80\%,~e^+: +30\%$).}
\label{fig:cmp_mh_fs}
\end{figure}

In order to validate the method, I compare the $M_{dl}$ and $M_{H}$ distributions of this fast simulation with the full simulation at $\sqrt{s}=$250~GeV. Figure~\ref{fig:mdl_mh_rdr250_fs} shows this comparison. They agree with each other. I have also done a further validation by repeating the analysis at $\sqrt{s}=$250~GeV for the ILD LOI based on this fast simulation, and comparing the results with those from the full simulation. The results are identical to each other.

Figure~\ref{fig:cmp_mh_fs} shows the comparison of the Higgs recoil mass distributions of the signal after fast simulation with various beam parameters considering the corresponding integrated luminosity.

\section{Analysis and Results}

The analysis procedure~\cite{thesis} after the fast simulation is similar to that of my study based on the full simulation samples:

\begin{itemize}
\item A cut based background suppression. The cuts are defined in Table~\ref{tab:cutchain}. These cuts are independent of the Higgs decay mode.
\item A Likelihood further rejection of background~\cite{thesis}. The variables ($P_{Tdl}$, $\cos{\theta}_{dl}$, $M_{dl}$ and $acolinearity$ ) employed in this Likelihood suppression are also independent of the Higgs decay mode.
\item The resulting recoil mass ($M_{recoil}$) spectrum of signal and background is fitted to derive the results of the $M_H$ and the $ZH$ cross-section measurement~\cite{thesis}. I choose the physics motivated function~\cite{thesis} to describe the signal. 
\end{itemize}

 \begin{table}[htbp] 
 \centering
    \begin{tabular}{l} 
       \hline 
       Cut-Chain \\ 
       \hline 
       (1)  $|\cos{\theta_{\mu}}|<0.99$\\
       (2) $P_{Tdl} > 20$~GeV \\
       (3) $M_{dl} \in (80,\ 100)$~GeV \\
       (4) $acop \in (0.2,\ 3.0)$ \\
       (8) $M_{recoil} \in (115,\ 150)$~GeV \\
       (9) Likelihood Further Rejection  \\
         \ \ \  (using variables $P_{Tdl}$, $\cos{\theta}_{dl}$, $M_{dl}$ and $acol$) \\
        \hline 
    \end{tabular} 
  \caption{The cut-chain for background suppression. } 
  \label{tab:cutchain}
\end{table}

After the background suppression, the remaining numbers of events of signal and background reactions are given in Table~\ref{tab:nevts}.

 \begin{table}[htbp] 
 \centering
    \begin{tabular}{c c c c} 
    \hline
    Reactions & $ZH\rightarrow\mu\mu X$ & $ZZ$ & $WW$ \\
    \hline
    $N_{initial}$ & 1248 & 29k & 61k \\
    $N_{selected}$ & 633 & 658 & 30 \\
    \hline
    \end{tabular} 
  \caption{Numbers of events before and after background suppression, for signal and backgrounds.} 
  \label{tab:nevts}
\end{table}

\begin{figure}[htbp]
\centering
\includegraphics[width=0.88\textwidth]{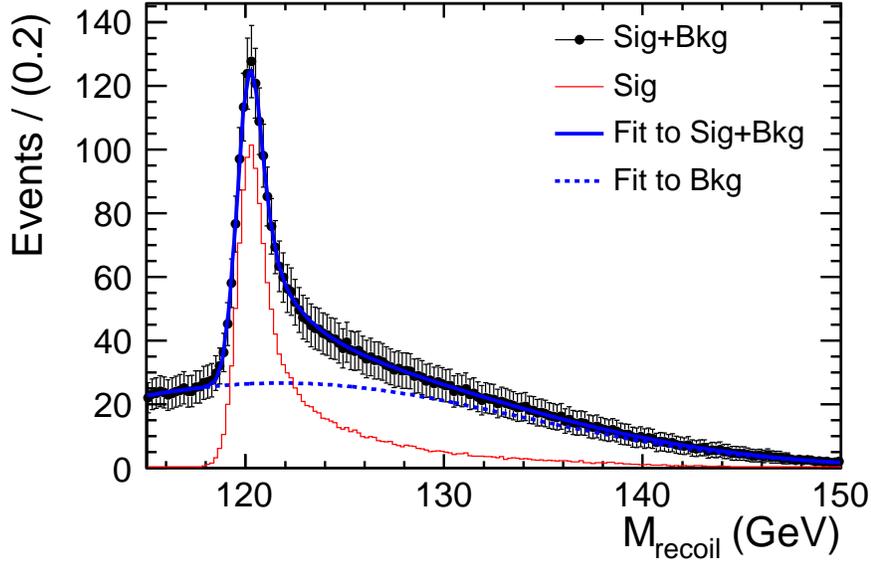}
\caption{Fit to recoil mass spectrum of signal plus background based on the fast simulation, at $\sqrt{s}=$350~GeV, with beam polarization ($e^-: -80\%,~e^+: +30\%$) beam parameters SB2009 w/o TF, and assuming an integrated luminosity of 175~fb$^{-1}$.}
\label{fig:fit_mi_lr_pm}
\end{figure}

 \begin{table}[htbp] 
 \centering
    \begin{tabular}{c|c c c c c } 
    Beam Par & $\mathcal{L_{\rm int}}$ (fb$^{-1}$) & $\epsilon$ & S/B & $M_{H}$ (GeV)& $\sigma$ (fb) ($\delta{\sigma}/\sigma$) \\ 
    \hline
    RDR 250 &			188	& 55\% & 62\% & $120.001\pm0.043$ & $11.63\pm0.45$  (3.9\%) \\
    RDR 350 &			300	& 51\% & 92\% & $120.010\pm0.084$ & $  7.13\pm0.28$  (4.0\%) \\
    SB2009 w/o TF 250b &	55	& 55\% & 62\% & $120.001\pm0.079$ & $11.63\pm0.83$  (7.2\%) \\
    SB2009 w/o TF 350 &	175	& 51\% & 92\% & $120.010\pm0.110$ & $  7.13\pm0.37$  (5.2\%) \\
    SB2009 TF 250b &	68	& 55\% & 62\% & $120.001\pm0.071$ & $11.63\pm0.75$  (6.4\%) \\
    SB2009 TF 350 &		250	& 51\% & 92\% & $120.010\pm0.092$ & $  7.13\pm0.31$  (4.3\%) \\
    \hline
    \end{tabular} 
  \caption{Results of different beam parameters, assuming a beam polarization of ($e^-: -80\%,~e^+: +30\%$). The results of RDR 250 and SB2009 w/o TF 250b are scaled from my previous analysis based on full simulation according to the integrated luminosity. That of RDR 350 is estimated by scaling the result of SB2009 w/o TF 350 obtained here according to the integrated luminosity.} 
  \label{tab:results}
\end{table}

Figure~\ref{fig:fit_mi_lr_pm} shows the fit to the recoil mass spectrum of remaining signal and background. An equivalent plot with beam parameters RDR 250 can be found in reference~\cite{thesis, note}. 

From Figure~\ref{fig:fit_mi_lr_pm}, the derived results of the Higgs mass measurement is $M_H=120.010\pm0.110$~GeV, and of the cross-section is $\sigma=7.13\pm0.37$~fb ($\delta{\sigma}/\sigma=5.2\%$). A comparison of the results with other beam parameters are shown in Table~\ref{tab:results}, together with the efficiency ($\epsilon$) and signal over background (S/B). In this table, the results of RDR 250 SB2009 w/o TF 250b are scaled from my previous analysis~\cite{thesis, note} based on full simulation according to the integrated luminosity. And the results of RDR 350 is estimated by scaling the result of SB2009 w/o TF 350 according to the integrated luminosity.
  
The higher S/B at $\sqrt{s}=$350~GeV is due to better background suppression. For example the variable $\cos{\theta}_{dl}$, its distribution of $ZH$ signal is much center for $\sqrt{s}=$350~GeV than 250~GeV~\cite{thesis}, while that of the $ZZ$ background is much forward.

When comparing RDR 250 and 350, the errors on the cross-section are similar, while the error on the $M_H$ at 350~GeV is worse by a factor of 2 than that at 250~GeV. 

When comparing SB2009 w/o TF 250 and 350, the higher luminosity at 350~GeV gives better result on cross-section, but not on the $M_H$, which is worse by 1.4 times due to the wider mass peak. 

For a given $\sqrt{s}$, the results of SB2009 w/o TF are worse due to the decrease of luminosity. The TF recovers the degradation to certain level.

Also, the comparison above shows the results on Higgs mass is about 3 times worse if we change to use beam parameters SB2009 350 w/o TF, and on the cross-section more than 1.5 times worse.

\end{document}